\documentstyle[12pt,epsf,epsfig]{article}
\def\beq{\begin{equation}}
\def\eeq{\end{equation}}
\def\bea{\begin{eqnarray}}
\def\eea{\end{eqnarray}}
\def\bq{\begin{quote}}
\def\eq{\end{quote}}

\def\NP{{\it Nucl.Phys.} }
\def\PL{{\it Phys.Lett.} }
\def\PR{{\it Phys.Rev.} }
\def\PRL{{\it Phys.Rev.Lett.} }

\def\hllap{h\llap {$/$}}

\def\gappeq{\mathrel{\rlap {\raise.5ex\hbox{$>$}}
{\lower.5ex\hbox{$\sim$}}}}

\def\lappeq{\mathrel{\rlap{\raise.5ex\hbox{$<$}}
{\lower.5ex\hbox{$\sim$}}}}

\def\Toprel#1\over#2{\mathrel{\mathop{#2}\limits^{#1}}}

\begin{document}
\pagestyle{empty}
\begin{flushright}
{CERN-TH/2002-226}\\
%hep-ph/th number??\\
\end{flushright}
\vspace*{5mm}
\begin{center}
{\bf HISTORY OF SPIN AND STATISTICS} \\
\vspace*{1cm}
{\bf A. Martin} \\
\vspace{0.3cm}
Theoretical Physics Division, CERN \\
CH - 1211 Geneva 23 \\
and\\
LAPP, Chemin de Bellevue, F-74941 Annecy-le-Vieux\\
\vspace*{2cm}
{\bf Abstract} \\ \end{center}
\vspace*{1cm}
\noindent
These lectures were given in the framework of the ``Dixi\`eme s\'eminaire 
rhodanien de physique'' entitled ``Le spin en physique'', given at Villa 
Gualino, Turin, March 2002.  We have shown how the difficulties of 
interpretation of atomic spectra led to the Pauli exclusion principle and 
to the notion of spin, and then described the following steps:  the Pauli 
spin with 2$\times$2 matrices after the birth of ``new" quantum mechanics, 
the Dirac equation and the magnetic moment of the electron, the spins and 
magnetic moments of other particles, proton, neutron and hyperons.  
Finally, we show the crucial role of statistics in the stability of the 
world.

 \vspace*{5cm}
%\noindent
%\rule[.1in]{16.5cm}{.002in}

\vspace*{0.5cm}

\begin{flushleft} CERN-TH/2002-226 \\
August 2002
\end{flushleft}
\vfill\eject
\pagestyle{empty}
\clearpage\mbox{}\clearpage

\setcounter{page}{1}
\pagestyle{plain}

%INSERT YOUR TEXT HERE
\underline{BIBLIOGRAPHY}

\underline{I. Most essential references}
\begin{itemize}
\item S.I. Tomonaga, The Story of Spin, English translation by T.Oka, The
University of Chicago Press, 1997.
\item Theoretical Physics in the 20th Century, Pauli memorial volume, M.
Fierz, V.F. Weisskopf eds., Interscience 1959.
\item Max Born, Atomic Physics, 7th edition 1962, Blackie and Son, Glasgow.
\end{itemize}

\underline{II. Additional Material}
\begin{itemize}
\item E. Chpolski, Physique Atomique, traduction fran\c caise, Editions Mir
1978.
\item F. Gesztesy, H. Grosse and B. Thaller, Phys. Rev. Letters 50 (1983)
625.
\item W. Thirring, Lehrbuch der Mathematische Physik, Quanten Mechanik
Grosser systeme (Springer 1980).
\item H. Grosse and A. Martin, Particle Physics and the Schr\"odinger
Equation, Cambridge University Press, 1997.

\end{itemize}
\vfill\eject

\section{Introduction}

Before speaking of spin, I have to speak of angular momentum, and
specifically classical angular momentum.
We know since a long time that angular momentum is a conserved quantity.
Specifically one of Kepler's laws of the
motion of planets around the sun, the ``law of areas'' is nothing but the
conservation of angular momentum:

If a planet moves around the sun, the time taken by the planet to go from 1
to 2 is equal to the time
taken to go from 3 to 4 if the shaded areas, delimited by the trajectory and
rays coming from the sun, are equal.

\begin{figure}[h]
\hglue4.5cm
\epsfig{figure=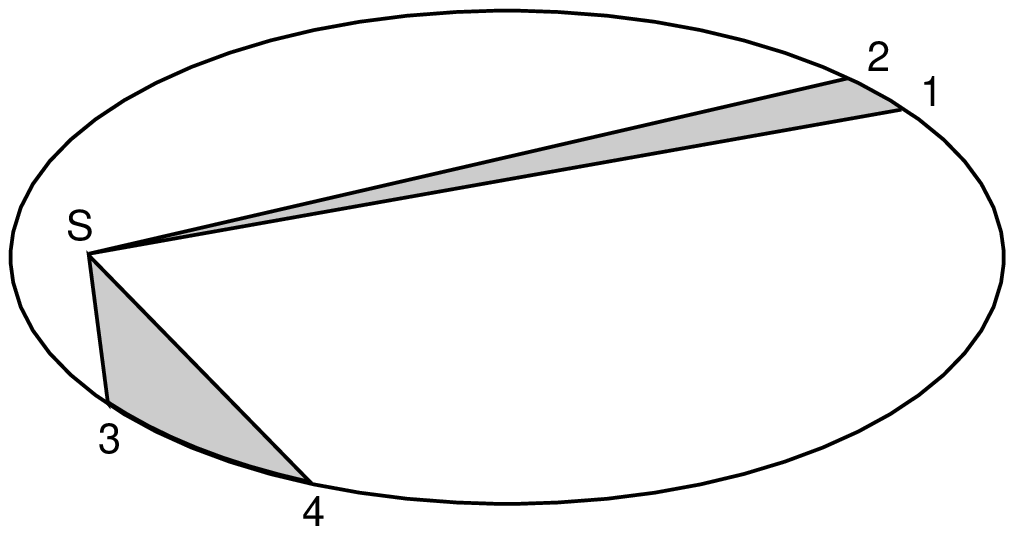,width=8cm}
\begin{center} Figure 1 \end{center}
\end{figure}

The angular momentum of the planet is
\beq
\vec x \Lambda \vec p = m \vec x \Lambda \frac{d \vec x}{dt}~,
\label{one}
\eeq
its derivative with respect to time is
\beq
m~\frac{d}{dt} \left[\vec x \Lambda \frac{d\vec x}{dt}\right] =  m~\frac{d
\vec x}{dt} \Lambda
\frac{d \vec x}{dt} + \vec x \Lambda \frac{d^2\vec x}{dt^2}~m~.
\label{two}
\eeq
The first term is obviously zero.  From Newton's law for a central force
\beq
m~\frac{d^2 \vec x}{dt^2} = \frac{\vec x}{r} F(r)~, {\rm where}~ r = |\vec
x|~,
\label{3}
\eeq
and hence the second term in (\ref{two}) is also zero.  The angular momentum
is a constant of motion,
and this \underline{ does not depend on the force behaving like $r^{-2}$}.
However, $\vec x \Lambda d\vec x$
represents precisely twice the area spanned by the rays going from the sun 
to the
planet during an interval $dt$.
 Hence the
``area law'' holds at the infinitesimal level.

Since Bohr's model of the atom (1913) angular momentum was quantized
\beq
|\vec L| = \ell  h \llap {$/$}~,
\label{four}
\eeq
and, in addition, the projection of the angular momentum along the $z$ axis,
$M$ was also quantized:
\beq
M = m  h \llap {$/$}~.
\label{five}
\eeq
The energy levels of atoms, especially \underline{Alcaline} atoms (the first
row of Mendeleieff's classification)
 were characterized by $n$ (principal quantum number) and $\ell$ (orbital
angular momentum, called $k$ by the pioneers
-- but I prefer to use the modern notation).

However, take the yellow light of Sodium, that light which illuminates
tunnels and makes you look like a corpse.
With a rather ordinary spectrometer that first year students were using at
the Ecole Normale Sup\'erieure when I was
there, you discover that this $3P \rightarrow 2S$ line ($n = 3, \ell = 1$ to
$n = 2, \ell = 0$) is in fact a doublet, the
relative spacing between the two lines being about 1/1000.  This means that
there is something wrong with the
classification of energy levels and spectral lines.

\section{The Pre-Spin, Pre-Quantum Mechanics Periods and the Discovery of
Spin}

In first approximation, in the Bohr model, the energy levels of Hydrogen
depended only one quantum number
the ``Principal quantum number'', $n$, as indicated on Fig.~2

\begin{figure}[h]
\hglue4.5cm
\epsfig{figure=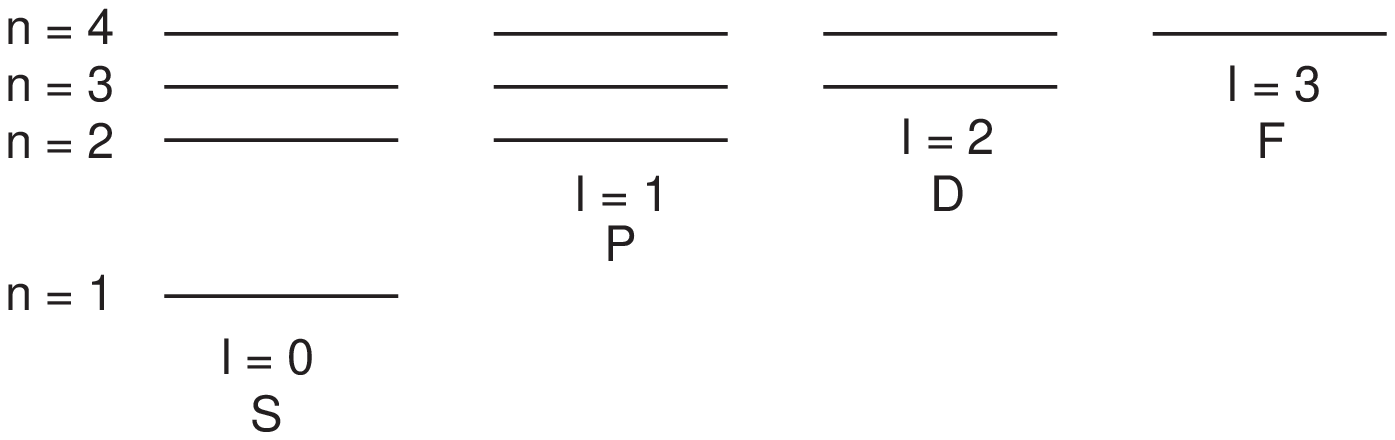,width=8cm}
\begin{center} Figure 2 \end{center}
\end{figure}

The energy levels were degenerate in $\ell$, the orbital angular momentum.

In Alcaline atoms, with a single valence electron, the energy levels also
depend on $\ell$, in such a way

\begin{figure}[h]
\hglue4.5cm
\epsfig{figure=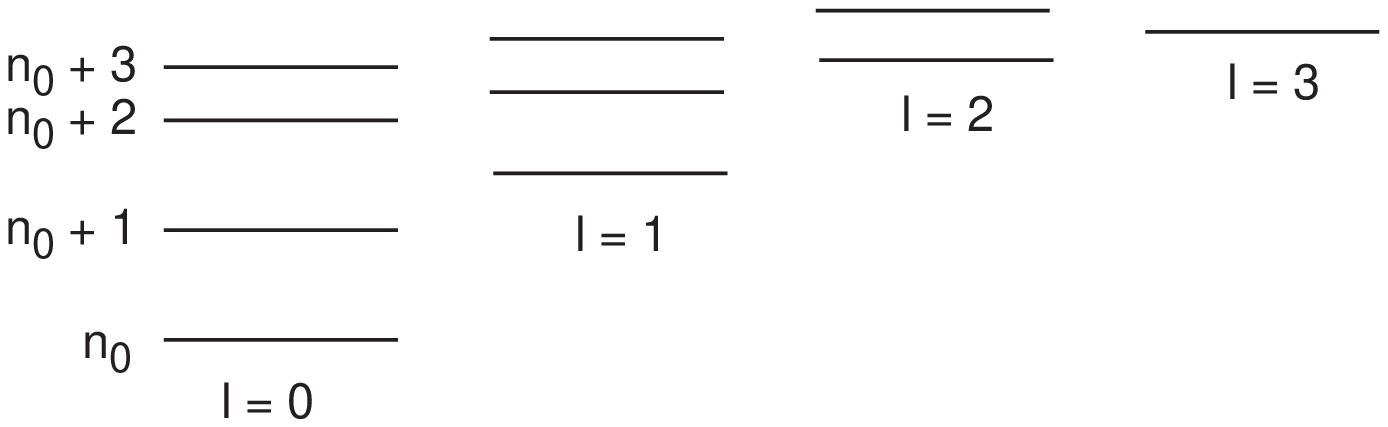,width=8cm}
\begin{center} Figure 3 \end{center}
\end{figure}

\noindent that for a fixed principal quantum number, the energy
\underline{increases} as $\ell$ \underline{increases}.
This is outside the subject but let me say that the best, most rational
explanation of this fact was only given
relatively recently by Baumgartner, Grosse and myself \cite{aaa}.  It is due
to the fact that the valence electron is
submitted to the field of the nucleus which is pointlike at this scale and
of the electron cloud which has a negative
charge and hence produces a potential with negative laplacian.

The observed transitions between levels satisfy the rule $|\Delta \ell| =
1$.
 However, in fact, as we already mentioned it, the levels of the Alcaline
atoms are doublets.  Similarly, the levels
of the Alcaline earths, with Magnesium as a prototype, with \underline{two}
valence electrons are either singlets or
triplets.  This is illustrated on Fig. 4.

\begin{figure}[h]
\hglue4.5cm
\epsfig{figure=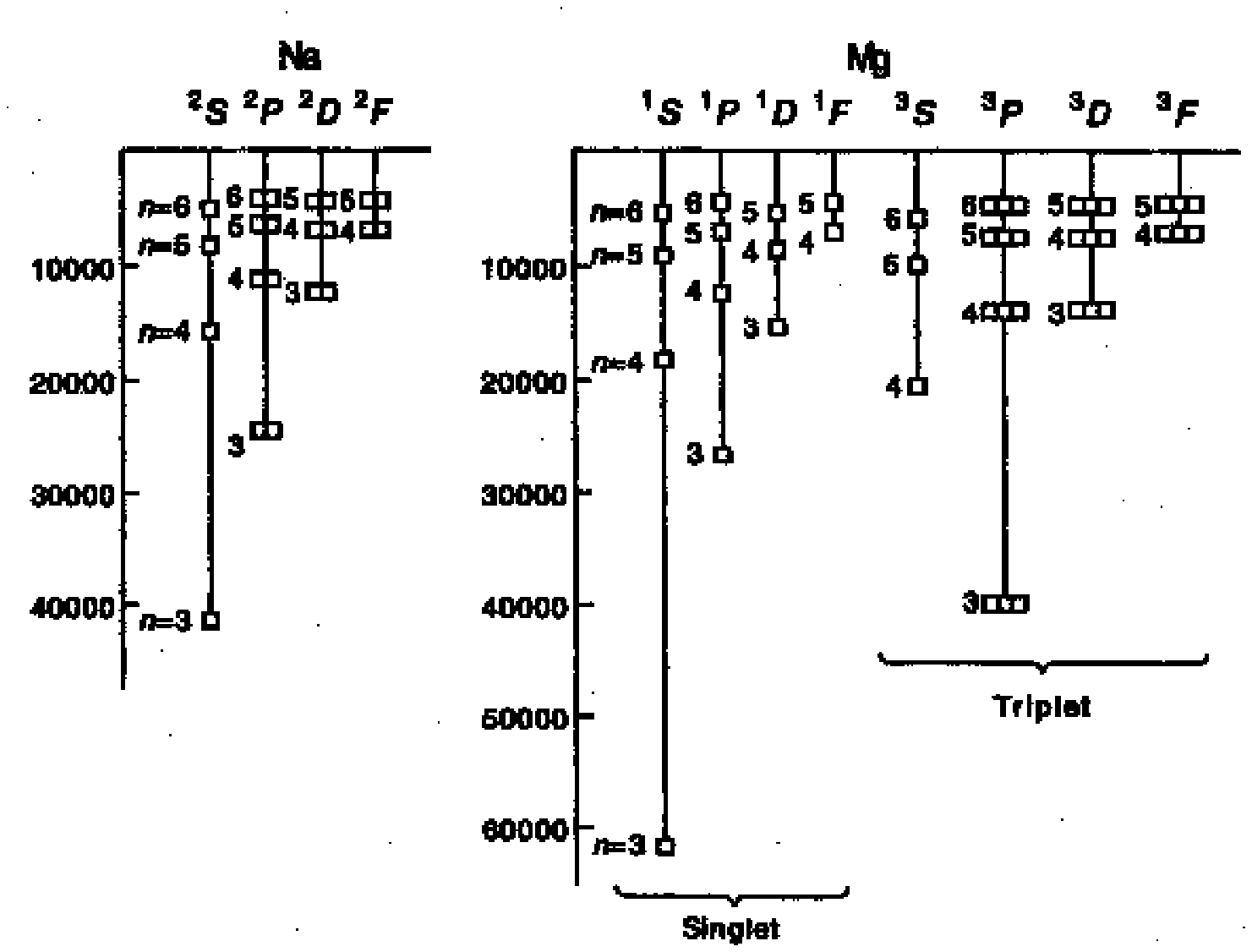,width=8cm}
\begin{center} Figure 4 \end{center}
\end{figure}

Another important piece of the puzzle is the Zeeman effect:  the splitting
of
the energy levels inside a uniform magnetic field.  In a \underline{weak}
magnetic field, one observes, in the case of
Alcaline atoms, that except for the $\ell = 0$ state which is a singlet, all
other doublets, for $\ell \not= 0$, split
in such a way that one member gives $2\ell$ levels and the other $2\ell + 2$
levels.  This could not be explained by
what was known at the time.

What could be explained was the so-called ``NORMAL'' Zeeman effect,
according to which a
level with orbital angular momentum component $m$ along the $z$ axis would
be shifted by a magnetic field $H$ in the
$z$ direction by
\beq
\Delta E (n,m,\ell) = \frac{e h \llap{$/$}}{2mc}~Hm~,
\label{six}
\eeq
$\frac{e h \llap{$/$}}{2mc}$ is called the Bohr magneton.

This is because an electron circulating around the atom is equivalent to a
current producing
a magnetic moment.

Similarly, in strong magnetic fields (the so-called Paschen-Back effect),
what is seen is
also very different.  We also see more levels than we should. Both
situations are illustrated on Fig.~5.

\begin{figure}[h]
\hglue4.5cm
\epsfig{figure=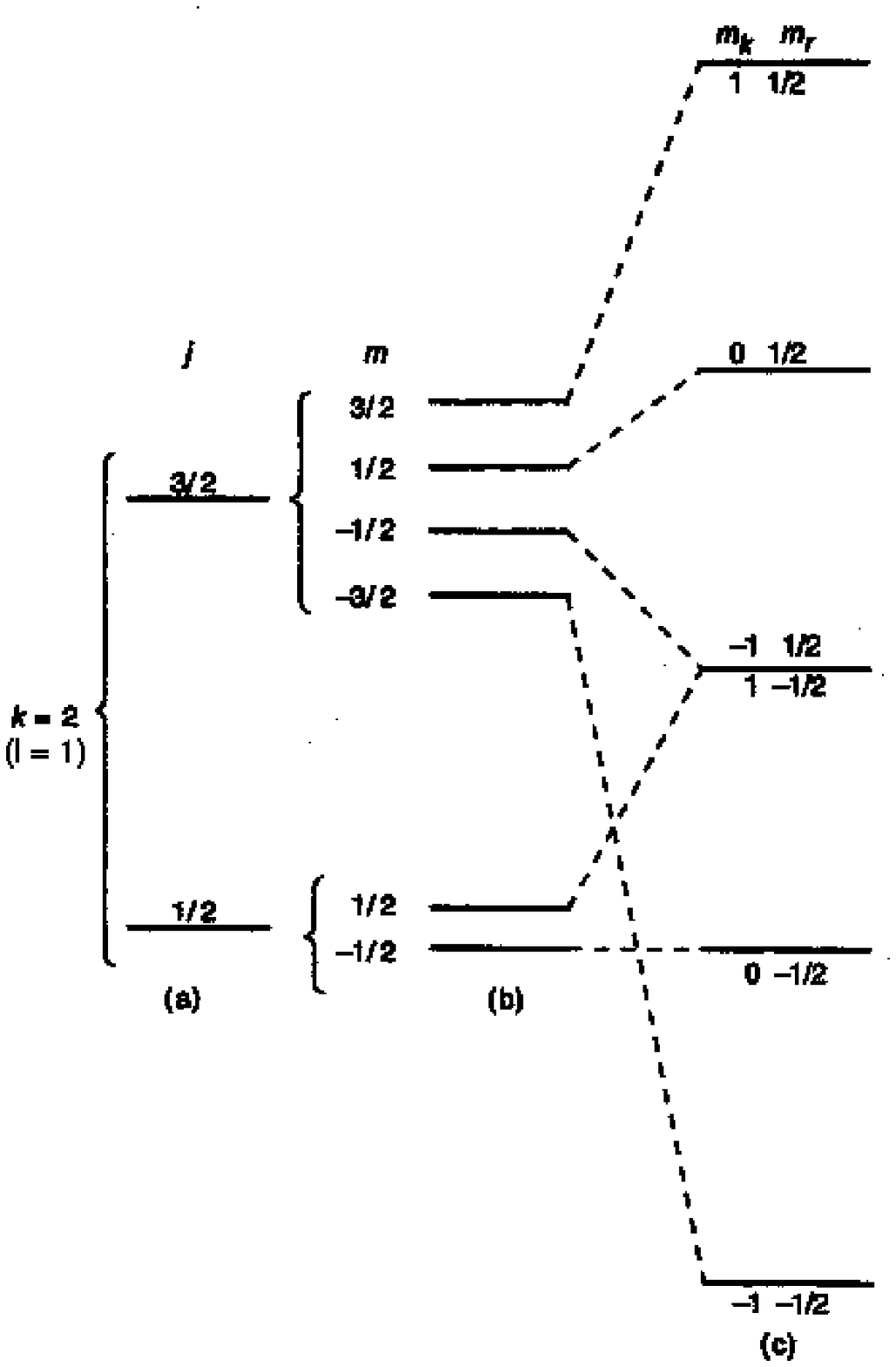,width=8cm}
\begin{center} Figure 5 \end{center}
\end{figure}

It is of course difficult to describe the difficulties of the pioneers
seeing this, because
we are somehow like in a TV episode where we know already who is guilty,
but inspector Columbo does not know yet!

Among the main protagonists were Land\'e, Pauli and Sommerfeld.  All of them
realized that an
extra quantum number beside $n, \ell$ and $m$ was needed.  Their models
fitted the data, but they knew they were only
models.  They were called ``ERSATZ'' models.  Those of you who have lived in
occupied Europe during the war know what
an ersatz is. It is a bad substitute for something much better.  However, in
this case, the ERSATZ was replacing
something which did not exist yet.

Sommerfeld and Land\'e were putting the fourth quantum number in the
``core'' of the atom
without a very precise definition of the core (nucleus?  or nucleus plus
inner electrons?).  The core was carrying
angular momentum, 1/2 in the case of the alcaline atoms, had a magnetic
moment, twice what you get in the case orbital
angular momentum:  2 $\times~ 1/2~ \times$ electron Bohr magneton. That was in
1923.

\underline{Pauli did not like that}, for many reasons.  For instance, if you
remove an electron from $Mg$,
you get $Mg^+$ which behaves like an Alcaline atom, except for the fact that
it is not neutral.  By miracle, the
``core'' which had angular momentum 0 or 1 suddenly gets angular momentum
1/2.

Pauli, in early 1925, decided that an extra quantum number was a ``classically
undescribable two-valuedness"
which belonged to the \underline{electron}, and formulated his famous
\underline{exclusive principle}, according to
which two electrons cannot be in the same state.  With this principle you
``explain'' the filling of the successive
shells.  For instance, sodium is

\begin{figure}[h]
\hglue4.5cm
\epsfig{figure=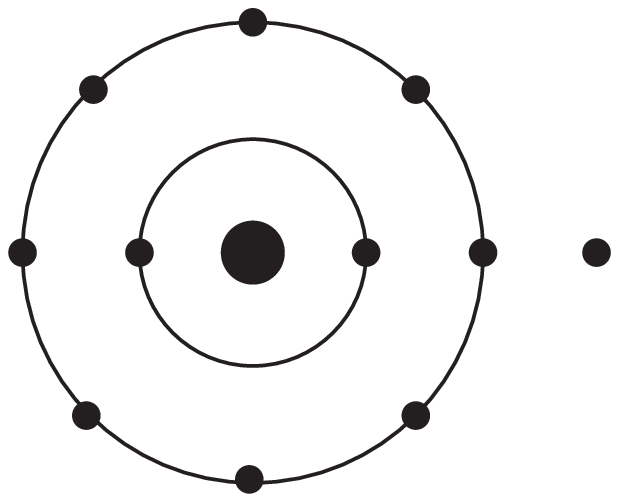,width=8cm}
\begin{center} Figure 6 \end{center}
\end{figure}

You explain also the counting of levels in weak and strong magnetic fields
as  it is seen in Fig.~5.

Pauli was not, however, willing to make the big jump that this is due to an
intrinsic angular
momentum $1/2~h \llap {$/$}$ of the electron.  It was R.L. Kronig, arriving
in Germany from the U.S. who proposed this
idea first.  However, this idea was not well received by Pauli, as well as
in Copenhagen where Kronig went visiting.
There was also a problem about the spacing of the levels which gave doubts
to Kronig himself.  Then in the fall of
1925, Uhlenbeck and Goudsmit, in Leiden, proposed the same idea which they
sent for publication to Naturwissenshaften.
After discussions with Lorentz, they tried to withdraw their paper, but it
was too late (fortunately) and it was
published!

Naturally the gyromagnetic factor of the electron was taken to be 2 to fit
the experiment,
so that the magnetic moment of the electron was $2 \times 1/2 \times 1$ Bohr
magneton.

There was a problem, however, which is that a na\"\i ve calculation of the
energy interval
in a doublet, due to the coupling of the orbital motion with the spin of the
electron, gave a result \underline{twice
too big} as stressed by a letter of Kronig against the Uhlenbeck--Goudsmit
hypothesis.  It was G.H.~Thomas who realized
that relativistic effects in the motion of the electron should be treated
more carefully.  It amounts to replacing $g$,
the gyromagnetic factor of the electron, which was empirically 2, by $g-1$,
i.e. 1.  This is treated in the lectures
of E.~Leader and so I shall not attempt to describe the reasoning of Thomas.

Before going to the true quantum mechanical situation, I would like to show
to you how, in the time of ``old'' quantum mechanics, people calculated the
``Land\'e factor'' giving the magnetic splittings in weak fields.  The
magnetic interaction is given by
\beq
W = - \frac{e h \llap {$/$}}{2mc} \langle H_{\mu_{''}} \rangle
\label{seven}
\eeq
where $\mu_{''}$ is the projection of the total magnetic moment on the total
angular momentum
$\vec L + \vec S = \vec J$;
\beq
\mu_{''} = \left( \frac{\vec J \cdot \vec\mu}{J^2} \right) \vec J
\label{eight}
\eeq
and $\vec\mu = \vec L + g_0\vec S$ with $g_0 = 2$.  So
\bea
\langle H \mu_{''} \rangle &=& - \left[ 1 + (g_0-1) \frac{\vec J \vec
S}{J^2}\right](\vec H \cdot \vec  J)  \nonumber \\
\langle H \mu_{''}\rangle &=& -g (H \cdot J)~,
\label{nine}
\eea
where $g$ is Land\'e factor.
\beq
g = 1 + (g_0 -1) \frac{J^2 + S^2 - L^2}{2J^2}~.
\label{ten}
\eeq
If you put the actual values of $j, S( = 1/2)$, and $\ell$, it does not
work!
You should use the substitution rule
\beq
J^2 \rightarrow j(j+1)~,~~S^2 \rightarrow s(s+1)~,~~L^2 \rightarrow
\ell(\ell+1)~,
\label{eleven}
\eeq
that I learnt from my old master Alfred Kastler who, in spite of the fact
that he
was a great physicist (he got the Nobel prize for ``optical pumping'' a
method to polarize atoms by sending circularly
polarized light on them), never took the pain to learn modern quantum
mechanics.

The fact is that this substitution rule gives the \underline{correct}
result:
\beq
g = \frac{3}{2} + \frac{s(s+1)-\ell(\ell +1)}{2j(j+1)}
\label{twelve}
\eeq
To me this was not obvious, and I checked it by hand!

Finally, I would like to speak of something which historically took place
\underline{before}
the critical years of the discovery of spin and \underline{should} have
accelerated things,
 namely the Stern--Gerlach
experiment (1921), which consists

\begin{figure}[h]
\hglue4.5cm
\epsfig{figure=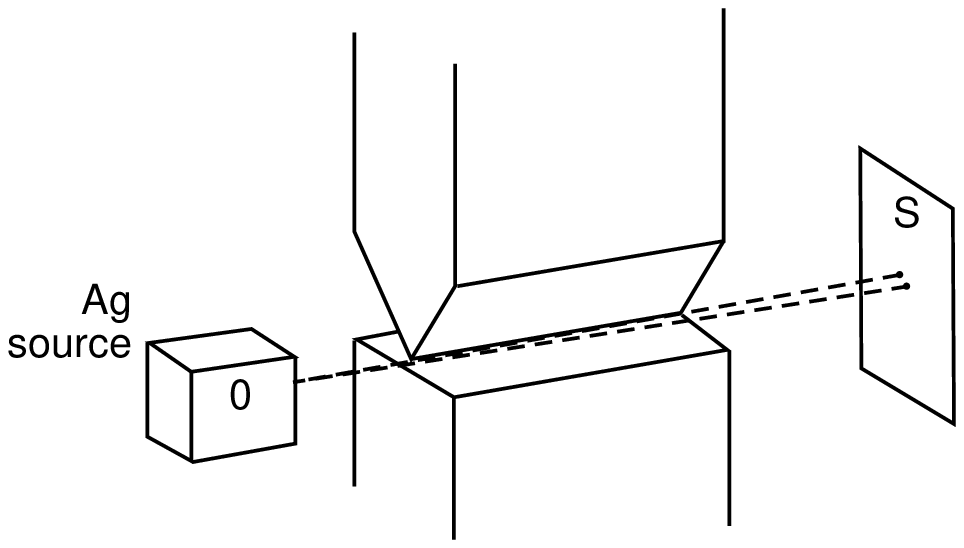,width=8cm}
\begin{center} Figure 7 \end{center}
\end{figure}

\noindent in sending a beam of silver atom through an
\underline{inhomogeneous} magnetic field.
The atoms behave like tiny magnetic dipoles and depending on their
orientation are deviated upward or downward.  At
the end one observes \underline{two} dots of silver atoms \underline{and
nothing else} (Fig.~7).  Remembering that $Ag$ is
monovalent, this fits with the spin picture.  However, this experiment seems
to have had little influence on the mind
of the great theoreticians in 1921.  This experiment is
\underline{fundamental} from another point of view.  It shows
that spin is \underline{absolutely not classical}.  You expect that $Ag$
atoms, in the source, have their spins
oriented at RANDOM.  Yet there are two and only two dots at the other end.  
This experiment, besides proving that the silver atom carry magnetic moment
with two possible orientations,
is also absolutely fundamental because it leads to a complete revision of
our ideas on what things are observables and
how they can be observed.  I shall come back on this later.

\section{Spin after the birth of ``modern'' quantum mechanics}

When new quantum theory was born?  In 1925 with Heisenberg's matrix
mechanics or in fact
\underline{before}, with the famous equation of Louis de Broglie (1924)
giving the wave length associated to a
particle:
\beq
\lambda = \frac{h}{p} = \frac{h}{mv}~~{\rm (in~the~non-relativistic~case)}~.
\label{thirteen}
\eeq
Anyway, shortly after the work of Heisenberg, Pauli at the end of 1925,
succeeded in getting
the energy levels of Hydrogen, by purely algebraic methods, using matrix
mechanics.

However, the news of De Broglie's relation (which Langevin had agreed to
present at the French
Academy of Sciences after consulting Einstein) was carried to Z\"urich by a
chemist named Victor~Henry (information
from David Speiser, son in law of Hermann Weyl).  Debye then told
Schr\"odinger: ``you should find an equation for
these waves"'.  He did, in 1926!  Schr\"odinger and others proved the
equivalence of the Schr\"odinger approach and of
Heisenberg's approach.  Schr\"odinger too found the energy levels of
hydrogen using properties of ``special
functions'".

The Schr\"odinger equation (with units such that \hllap =1) looks like:
$$
-\frac{1}{2m} \Delta \psi + (V-E)\psi = 0~,
$$
which, for a \underline{central} potential in polar co-ordinates gives
\bea
&&-\frac{1}{2m}~\frac{1}{r^2}~\frac{d}{dr}~r^2~\frac{d}{dr}~\psi \nonumber
\\
 && +\frac{1}{2mr^2} \times
%\mbox{
(-1) \left[ \frac{1}{\sin \theta} \frac{\partial}{\partial \theta} \left( \sin
\theta
\frac{\partial}{\partial\theta} \right) +
\frac{1}{\sin^2\theta}~\frac{\partial^2}{\partial \phi^2}\right] \psi
\nonumber \\ && + (V-E)\psi = 0~.
\label{fourteen}
\eea
It is tedious but trivial to check that the second line in Eq.~(\ref{fourteen}) is nothing but the
square of the angular momentum operator acting on $\psi$:
\beq
L^2 = L^2_x + L^2_y + L^2_z~,
\label{fifteen}
\eeq
\beq
\vec L = \vec x \wedge \vec p~.
\label{sixteen}
\eeq
On the one hand $p$ admits the representation
\beq
p_x = -i \frac{\partial}{\partial x}~,~~p_y = -i \frac{\partial}{\partial
y}~,~~ p_z = -i
\frac{\partial}{\partial z}~.
\label{seventeen}
\eeq
On the other hand, it has the commutation relation
\bea
\left.
\matrix{
&&[p_x,x] = -i  \hfill\cr
&&[p_x,y] = 0 \hfill\cr
&& etc.\hfill\cr
&&{\rm (\times~\hllap~in~general~units)} } \right\}
\label{eighteen}
\eea
At that point, one can take \underline{two} attitudes.
\begin{itemize}
\item[i)] Use the fact that the Schr\"odinger equation is separable for a
central potential
and look for the eigenfunctions and eigenvalues of $L^2$, represented by the
differential operator on the second line of Eq.~(\ref{fourteen}).  To do this it suffices to open a good book on special functions
(for instance, the ``Bateman" Project).  One
finds that the eigenvalue of $L^2$ are $\ell(\ell+1)$ where $\ell$ is an
\underline{integer} and the eigenfunctions
are the spherical harmonics
$Y_{\ell m}(\theta,\phi)$

\beq
Y_{\ell,m}(\theta,\varphi) = C(\sin \theta)^m \left(\frac{d}{d
\cos\theta}\right)^mP_\ell
(\cos\theta)e^{im\phi}~,
\label{nineteen}
\eeq
the $P_\ell$'s being Legendre polynomials.

\item[ii)] The second attitude consists in using the algebraic properties of
the operators $L_x, L_y, L_z$ and in particular their commutation relations:
\bea
\left.
\matrix{
L_x L_y - L_y L_x &=& iL_z \cr
L_yL_z - L_zl_y &=& il_x \cr
L_zL_x - L_xL_z &=& iL_y }\right\}
\label{twenty}
\eea
from which it follows that
\beq
[L_z,L^2] = 0, etc...
\label{twentyone}
\eeq
You start from an eigenstate of $L^2$ and $L_z$.  Say that
\beq
L^2\psi = \ell(\ell +1) \psi~.
\label{twentytwo}
\eeq
$\ell$, so far, is anything real ($L^2$ is a Hermitean operator!)
\beq
L_z\psi = m \psi~.
\label{twentythree}
\eeq
\end{itemize}

From positivity of $L^2_x, L^2_y, L^2_z$, it is already clear that
$$
| m | \langle \sqrt{\ell(\ell+1)}~.
$$
Now it is easy to see that the operators
\bea
\left.
\matrix{
L_+ &=& L_x + iL_y \cr
{\rm and}\hfill\cr
L_- &=& L_x - iL_y} \right\}~,
\label{twentyfour}
\eea
applied to $\psi$ raise or increase $L_z$ by \underline{one unit} without
changing $L^2$.
It is also easy to see that these operators must terminate both ways and
that $m_{\rm maximum} = \ell, m_{\rm
minimum} = -\ell$, therefore $2\ell +1$ is \underline{integer}, and indeed it is
acceptable to have \underline{integer}
orbital momentum as in the Schr\"odinger equation.  However, there is the
possibility, if we use only the
algebraic structure of angular momentum, to have \underline{half integer}
angular momentum.

The group theoretical interpretation of the  conservation of angular
momentum is that physics,
in the case of a central potential, is invariant under rotations around the
origin.  $L_x, L_y, L_z$ are the
generators of the rotation group around respectively the $x$ axis, the $y$
axis, the $z$ axis. A rotation around
$L_x$ of angle $\theta_x$ is
$$
{\rm exp} ~i \theta_x L_x~.
$$
A general eigenstate of $L^2$, with eigenvalue $\ell(\ell +1)$, can be
represented as a vector
column with components corresponding to $L_z = -\ell, -\ell +1, \cdots,
+\ell$.  Acting with $\exp i
\theta_xL_x$ on this column will produce a new state again with $L^2 =
\ell(\ell +1)$.  Clearly, the action of
$\exp ~i \theta_xL_x$ will be represented by a matrix with $(2\ell + 1)
\times (2\ell +1)$ components, i.e., we
generate in this way a \underline{representation} of the rotation group of
dimension $2\ell +1$.

Now, take the simple case of $\exp i \theta_zL_z$, acting on a column made
of components which
 are eigenstates of $L_z$.  If $\theta_z = 2\pi$, we come back to the
initial state, i.e., we should find the
\underline{same} vector we started from.  A component with $L_z = m$ will be
multiplied by $\exp 2i \pi m$,
which  will be \underline{unity} if $\ell$ is integer.  So, to represent the
true rotation group, we must use
only integer $\ell$ and $m$.

However, there is another group, which is the covering group of the rotation
group, and which
is in fact $SU_2$, the group of $2 \times 2$ unitary transformations, which
has the same \underline{Lie
algebra}, i.e., the \underline{same} commutation relations between $L_x,
L_y$ and $L_z$ the generators of the
group, but which admits both \underline{integer values} of $\ell$ and
\underline{half integer values} of
$\ell$.  In the case of an half integer $\ell$, a column vector will be
multiplied by -1 in a complete turn. The
half integer angular momenta will be also acceptable if this -1 factor is
invisible in physical results, which
is the case.

In what follows, $\ell$ will be restricted to designate the
\underline{orbital} angular momentum,
always integer.  From what we have said, an angular momentum 1/2 for the
electron is acceptable, and this is
precisely what Pauli exploited.  He understood that it was pointless to try
to represent the spin wave function
in $x$ space, and it should be in fact represented as a two-component
column, $ \left(\matrix{  \alpha \cr \beta
}\right)$ with $\alpha^2 + \beta^2 = 1$, $\left( \matrix{ 1 \cr 0 }\right)$
being a state with total angular
momentum 1/2 and projection on the $z$ axis +1/2, while $\left( \matrix{ 0
\cr 1} \right)$ has a projection on
the $z$ which is -1/2.

The spin operators satisfy the same algebra as the compoments of $L$:
\bea
S_xS_y - S_y S_x &=& iS_z \nonumber \\
S_y S_z - S_zS_y &=& iS_x \nonumber \\
S_z S_x - S_xS_z &=& iS_z
\label{twentyfive}
\eea
and $\vec S^2 = 3/4$.

They can be represented very simply by the ``Pauli'' hermitian matrices:
\bea
\sigma _x = \left( \matrix{0 & 1 \cr 1 & 0 }\right)~,~~ \sigma_y =
\left( \matrix{0 & -i \cr i & 0 } \right)~, ~~ \sigma_z = \left( \matrix{ 1
& 0 \cr 0 & -1 } \right)
\label{twentysix}
\eea
with
\beq
S_x = {\sigma_x\over 2}~, \quad S_y = {\sigma_y\over 2}~, \quad S_z =
{\sigma_z \over 2}
\label{twentyseven}
\eeq
the $\sigma^\prime$s have two remarkable properties:
\bea
\left.
\matrix{
\sigma^2_\mu &=& 1 \cr
\sigma_\mu\sigma_\nu &=& -\sigma_\nu \sigma_\mu \quad{\rm if}\quad \nu \not=
\mu}\right\}
\label{twentyeight}
\eea
Now if we take into account:\\
- The postulate that the gyromagnetic factor of the electron is 2, based on
experiment \\
- The Thomas precession,\\
we get the Pauli Hamiltonian in a magnetic field $H$ associated to a
potential
vector $A$:
\beq
H_p = {(\vec p - e\vec A)^2\over 2m} + V(r) + {ie\over 2m} ~~ \vec\sigma
\cdot
\vec H + {1 \over 4m^2} ~~ {\vec\sigma\cdot\vec L\over 2} ~~^{``} {1\over
r}~~{dV\over dr} ~^{"}
\label{twentynine}
\eeq
The double quotes indicates that this Hamiltonian is not really a
Hamiltonian if
${1\over r}~{dV\over dr}$ is more singular than ${1\over r^2}$ near the
origin.
For a purely Coulombic potential ${1\over r}~{dV\over dr}$ behaves like
$r^{-3}$.
Then strictly speaking, $H_p$ is not lower bounded, which means that you can
find
a trial function which gives an expectation value of $H_p$ arbitrarily
negative.
There was a period when this difficulty was just disregarded because the
spin-orbit term was just regarded as a perturbation. Only relatively
recently
\cite{bb} \underline{as we shall see later}, it was understood that the spin
orbit
term \underline{must} be indeed treated as a perturbation, at least for a
one-electron system. Hence, the recipe is:\\
- Solve the Schr\"odinger equation without spin and without magnetic field
for a
given orbital angular momentum;\\
- Take the expectation value of ${1\over r}~{dV\over dr}$, which we denote
as
$<{1\over r}~{dV\over dr}>$;\\
- Then diagonalize the Hamiltonian which, for a constant magnetic field in
the
$Z$ direction and neglecting higher orders in $H$ is
\beq
H = {p^2\over 2m} + V(r) + {ie\over 2m} ~~(L_Z + \sigma_Z) ~H + {1\over
4m^2}~~{\vec L\cdot\vec\sigma\over 2}~~
\langle{1\over r}~{dV\over dr} \rangle
\label{thirty}
\eeq
$L^2$ commutes with $H$, and $J_Z = L_Z + {\sigma_Z\over 2}$ also commutes
with
$H$ since $\vec L\cdot\vec\sigma =  J^2 - L^2 - \left({\vec\sigma\over
2}\right)^2$.

So, for any magnetic field, all we have to do is to diagonalize a 2$\times$2
matrix. For instance, in the basis where $L_Z$ is diagonal we take as basis
vectors
\bea
&&\vert\ell , L_Z = J_Z + 1/2 >~~~ \vert S_Z = -1/2 > \nonumber \\
{\rm and}~~&&\vert \ell , L_Z =  J_Z -1/2 > ~~~\vert S_Z = +1/2 >~.
\label{thirtyone}
\eea

In the extreme case where
$${eH\over 2m} \ll {1\over 4m^2} < {1\over r}~{dV\over
dr}>$$
 we get the Zeeman effect where $J^2$ is a good quantum number and in the
case
$${eH\over 2m} \gg {1\over 4m^2} < {1\over r}~{dV\over
dr}> ~,$$
 we get the Paschen-Back effect (see Fig. 5).

Here I would like to introduce a disgression about the spin-orbit interation
and
the one-electron model. Of course, the one-electron model is good -- almost
perfect -- for hydrogen. For alcaline atoms, with B. Baumgartner, H. Grosse
and
J. Stubbe \cite{aaa},\cite{cc} we have obtained a lot of very stringent
inequalities on the
energy levels link to the fact that the potential $V(r)$, due to the nucleus
and
the electron cloud satisfies $r^2 \Delta V(r) < 0$ as long as spin is
disregarded
(i.e., if one takes averages over spin multiplets), these inequalities are
very
well satisfied. We have already mentioned
\beq
E(n,\ell ) < E(n, \ell +1)~,
\label{thirtytwo}
\eeq
which allows to understand why the third electron of lithium is in an $s$
state
and not a $p$ state. There are others linking three successive ``angular
excitations", i.e., levels with $n = \ell +1$, which work beautifully
\cite{cc}.

However, if one tries to go further and gets inequalities on the fine
splittings
\cite{dd} themselves, things do not always work. For levels such that $\ell
= n-1$,
define:
\beq
E(J = \ell +1/2) - E(J = \ell -1/2) = {2\ell+1\over 4m^2}~~\langle {1\over
r}~{dV\over dr}~\rangle = \delta (\ell )
\label{thirtythree}
\eeq
the first ``theorem" is that it is positive for $\Delta V < 0$. Furthermore,
denoting $E(\ell +1, \ell)$ as $E(\ell )$
\beq
\delta(\ell ) > {1\over m\ell}~~{(\ell+2)^4\over (2\ell +
3)^2}~~\bigg(E(\ell +1)
- E(\ell )\bigg)^2 = \Delta (\ell )
\label{thirtyfour}
\eeq
and
\beq
{\delta (2) \over \delta (1)} < {8\over 81}
\label{thirtyfive}
\eeq
These inequalities are almost satified by the LiI isoelectronic sequence of
ions,
and really satisfied starting from Carbon IV.

{\small
\vspace*{3.5cm}
\begin{center}
Table 1 \\
Fine splittings for the Li~I isoelectronic sequence in (cm)$^{-1}$. \\
In the one-electron model,
we should have\\
$\delta(1) > \Delta(1), \delta(2) > \Delta(2), \delta(2)/\delta(1) < 8/81$

\vglue0.3cm
\begin{tabular}{|l l l l l l l|}
\hline
&      & $\delta(1)$ & $\delta(2)$ & $\Delta(1)$ & $\Delta(2)$ &
$\delta(2)/\delta(1)$\\
\hline
Li & I     & ~~~~0.3372 & ~~~0.037 & ~~~~0.4218 & ~~0.0362 & 0.1097\\
Be & II    & ~~~~6.58   & ~~~0.55  & ~~~~6.8748 & ~~0.480  & 0.0836\\
B  & III   & ~~~34.1    & ~~~3.1   & ~~~34.287  & ~~2.540  & 0.0909\\
C  & IV    & ~~107.1    & ~~10.5   & ~~106.6    & ~~9.278  & 0.098\\
N  & V     & ~~258.7    & ~~22.0   & ~~256.54   & ~22.64   & 0.085\\
O  & VI    & ~~532.5    & ~~51.1   & ~~525.7    & ~46.9    & 0.096\\
F  & VII   & ~~975.8    & ~~90.0   & ~~926.66   & ~86.97   & 0.092\\
Ne & VIII  & ~1649.3    & ~150     & ~1638.4    & 148.3    & 0.091\\
Na & IX    & ~2650      & ~224     & ~2600      &          & 0.085\\
Mg & X     & ~3978      & ~470     & ~3938      & 362      & 0.118\\
Al & XI    & ~5800      & ~560     & ~5713      & 530      & 0.096\\
Si & XII   & ~8180      & ~990     & ~8112      & 752      & 0.121\\
P  & XIII  & 11310      & 1000     & 11100      &          & 0.088\\
S  & XIV   & 15130      & 1420     & 14900      &          & 0.094\\
Cl & XV    & 19770      & 1000     & 19560      &          & 0.051\\
Ar & XVI   & 25655      & 3600     & 253090     &          & 0.140\\
K  & XVII  & 32538      & 3040     &            &          & 0.093\\
Ca & XVIII & 40843      & 3810     & 40400      &          & 0.093\\
Sc & XIX   & 50654      & 4730     & 50200      &          & 0.094\\
Ti & XX    & 62143      & 5780     & 61500      &          & 0.093\\
V  & XXI   & 75499      & 7020     & 74800      &          & 0.093\\
Cr & XXII  & 90910      & 8440     & 90000      &          &\\
\hline
\end{tabular}
\end{center}}

However, if one looks at the Sodium isoelectronic sequence, one finds strong
violation of the inequalities, in particular $3D^{5/2} < 3D^{3/2}$, so that
even
the sign of the splitting is wrong.  So the spin splittings are more
sensitive
than the levels themselves to the detailed structure of the many-body wave
function.

Before going to the Dirac equation, I would like:
\begin{itemize}
\item[a)] to return to the Stern-Gerlach experiment that I have mentioned in
Section 2.
\end{itemize}

It allows not only to show that atoms, and in fact electrons, have a
magnetic moment, which can be measured in this way, but the fact that at the
end
of the apparatus, the silver jet is split and produces two and only two dots
aligned with  the direction of the magnetic field is of
\underline{fundamental
importance}. Suppose that silver atoms were just tiny \underline{classical}
dipole magnets. In the oven, via collisions, their orientation should be
randomly
distributed, and for instance a dipole oriented in the horizontal plane,
perpendicular to the direction of the magnetic field would not be deflected
vertically, neither up nor down. So, one would expect to see on the final
plate a
continuous segment. This, however, is not the case.

This means that the measuring apparatus projects the atoms on the
eigenstates of
$\sigma_Z$ if the magnetic field is in the direction $Z$, and  deflects then
because of the inhomogeneity. Yet the initial assembly of atoms can be as
well
seen as a statistical assembly of eigenstates of $\sigma_X$ or $\sigma_Y$.
The
measuring apparatus behaves like a projector of the states on the particular
eigenstates it prefers, whether we like it or not. Many very refined tests
of
quantum mechanics, as opposed to classical mechanics, have been performed
\cite{ee} but
this very old one stands and defies any classical explanation.
\begin{itemize}
\item[b)] to speak about the two-electron system, which is only the
beginning of
the \underline{many} electron systems.
\end{itemize}

After the birth of new quantum mechanics, you could not anymore state the
Pauli
principle by saying that two electrons are in a different state, because, if
you
accept that one can write a \underline{many-particle Schr\"odinger
equation}, with
\beq
H = -\sum_i ~~{1\over 2m_i}~~\Delta_i + V(x_i) + \sum_{i>j}~~W(r_{ij}) + {\rm
spin
~terms}
\label{thirtysix}
\eeq
the wave function $\psi(x_1, s_1, x_2, s_2, \dots, x_n,s_n)$ cannot be written as a
product. It seems that it is Heisenberg who understood that the ``new" Pauli
principle should be that the wave function should be completely
antisymmetric,
including the spin variables. If you switch off the interaction between
electrons, you could have a product wave function, but according to the new
rule,
it should be replaced by what is called a ``Slater determinant". If the
one-particle wave functions in the determinant are mutually orthogonal and
if
there is no interaction, the energy is the same as with the product wave
function, but as soon as the interaction between electrons is on, the Slater
determinant is infinitely superior and can be used as a trial wave function.
In
the special case of two electrons, the wave function is a product of the
co-ordinate wave function and of the spin wave function. Then the Pauli
principle
requires that\\
- if the space wave function is symmetrical the spin wave function should be
antisymmetrical;\\
- if the space wave function is antisymmetrical, the spin wave function
should be
symmetrical.

Things are extraordinarily simple because:

If the spin wave function is \underline{symmetrical} the total spin is ONE.
In
the special case where $\sigma_{z1} = \sigma_{z2} = 1$ the wave function is
just
\beq
\left({1\atop 0}\right)_1 \times \left({1\atop 0}\right)_2
\label{thirtyseven}
\eeq
which has $s = 1, s_Z = 1$ and which is \underline{symmetrical}.

Applying to this wave function, the lowering operator
$$
(\sigma_X - i\sigma_Y)_1 + (\sigma_X -i\sigma_Z)_2
$$
does not change the symmetry. Hence all spin 1 wave functions are
\underline{symmetrical}.

The $S = 1, S_Z = 0$ wave function is:
\beq
C \left[ \left({0\atop 1}\right)_1\left({1\atop 0}\right)_2 +
\left({1\atop 0}\right)_1+\left({0\atop 1}\right)_2~~\right]
\label{thirtyeight}
\eeq
so the wave function
\beq
C \left[ \left({0\atop 1}\right)_1\left({1\atop 0}\right)_2 -
\left({1\atop 0}\right)_1\left({0\atop 1}\right)_2~~\right]
\label{thirtynine}
\eeq
is orthogonal to it. It has necessarily spin 0, otherwise you would get $2
(S =
1~~~S_z = 0)$ wave functions which is impossible. Therefore the
antisymmetrical
wave function has necessarily spin zero. After the preliminaries, you can
embark
into atomic and molecular physics, and more generally quantum chemistry,
magnetism, etc., the first steps being the molecular hydrogen ion, the
Helium
atom, the hydrogen molecule.

Many numerical results have been obtained, some very accurate, but rigorous
results are still coming. It is only relatively recently that the stability
of
the hydrogen molecule has been rigorously established by J. Fr\"ohlich, J.M.
Richard, G.M. Graf and H. Seifert \cite{ff}, without using the
Born-Oppenheimer
approximation which is a two-step process:
\begin{itemize}
\item[a)] solving the Schr\"odinger equation for  fixed positions of the
protons;
\item[b)] using the electron energy plus the Coulomb repulsion between the
protons as potential between the protons.
\end{itemize}

Another application of the structure of the spin wave function of two
electrons
concerns Cooper pairs. At least in normal supraconductors it is known that
one
finds ``Cooper pairs", pairs of electrons bound by interaction with
the lattice which are in an $S$ state. Hence from the Pauli principle, their
spin
wave function is antisymmetric with spin zero. Suppose, like G.B. Lesovik,
T.
Martin and G. Blatter \cite{ggg}, that a Cooper pair crosses the border
between a
supraconductor and a normal conductor. The pair will dissociate but it will
``remember" that it has total spin zero, at least as long as  depolarizing
collisions are negligible. Then, if one measures the spins of the two
electrons,
one sees typical quantum effects.

Na\"\i vely, one would expect that if the spin of electron Nr. 1 is
\underline{up} in the $z$ direction, the spin of electron Nr. 2 is
necessarily \underline{down} in the $z$ direction.  This would be like to
what
happens with Bertlmann's socks \cite{hh} (R. Bertlmann is a slightly
excentric Viennese
physicist, friend of John Bell): if he has a red sock on the left foot you
can
predict that he has a blue sock on the right foot.

But quantum physics is not like Bertlmann. If, for instance, the spin of
electron 1 is found to be +1/2 in the $z$ direction, the spin of electron 2
can be
found to be + 1/2 in the $x$ direction. This is because the scalar product
$$
\left < \left({1\atop 0}\right)_1 \times \left( {1/ \sqrt{2}\atop
1/ \sqrt{2}} \right)_2  \bigg\vert {1\over\sqrt{2}}~~
\left( \left({1\atop 0}\right)_1~~\left({0 \atop 1}\right)_2 -
\left({1\atop 0}\right)_2~~\left({0 \atop 1}\right)_1  \right) \right >
$$
is not zero.

\begin{itemize}
\item[c)] to prepare the Dirac equation by pointing that already before
Dirac it
was known that  if in zeroth approximation the energy levels of hydrogen
depend
only on $n$, in first approximation they depend on $n$ and on $j$ the total
angular momentum, but not on $\ell$.
\end{itemize}
It is only since the discovery of the
Lamb Shift (1949)  that we know that there is a small dependence on the
orbital
angular momentum: $E (2S_{1/2} - 2 P_{1/2})$ = 1058 Megacycles.

We come now to the Dirac equation (1928). Paul Adrien Maurice Dirac was a
British
physicist whose father was coming from Switzerland, exactly Saint Maurice,
with
ancestors in Poitou who fled to Switzerland to escape forced recruitement in
Napoleon's army (story that Dirac told me.  He even said ''there are 
more famous people than me coming from Poitou, like Mr. Cadillac'').

Dirac was guided essentially by aesthetic reasons. He wanted to have a wave
equation \underline{linear} in $p$ and not only in $d/ dt$, because of the
relativistic links between space and time, which, for $A = V = 0$ would
reduce to
the Klein-Gordon equation which gives the energy of a freely moving particle
\beq
(p^2 + m^2 - E^2)~\psi = 0
\label{forty}
\eeq
There was no possible solution using ordinary numbers, but Dirac was
strongly
influenced by Pauli's recent work using matrices. So he tried
\beq
H_0 = \sum^3_{i=1}~\alpha_ip_i + \beta m~,
\label{fortyone}
\eeq
and found that, to reproduce the Klein-Gordon equation you need
$$
\matrix{
\alpha^2_i = 1 &   \alpha_i\alpha_j
  + \alpha_j\alpha_i = 0~~{\rm if}~Z~ i\not= j \cr
\beta^2 = 1 & \alpha_i\beta + \beta \alpha_i = 0\hfill}
$$

The solution consists in taking the $\alpha_i$'s and $\beta$ to be 4$\times$
4
matrices.

It was proved by Van Der Waerden \cite{jj} that the only other solutions
are just
repetitions of 4$\times$ 4 matrices, modulo a change of
basis.

The $\alpha$ matrices can be written in terms of 2$\times$ 2 matrices. Then,
using the notation of H. Grosse one has
$$
H - mc^2 = \left|
\matrix{V &\vcenter{\vbox to 25pt{\leaders\vbox to
4pt{\vfil\hbox{.}\vfil}\vfil}} & c \vec \sigma \cdot \vec \pi\cr
%\noalign{\vskip -10pt}
\multispan3{\dotfill}\cr
c\vec \sigma\cdot \vec \pi & \vbox to 20pt{\leaders\vbox to 4pt{\vfil\hbox{.}\vfil}\vfil}&  V - 2mc^2\cr}\right|
$$
with $\vec\pi = - i ~h \llap {$/$} \vec\nabla - e \vec A$, each subsquare
being a ~ 2$\times$ 2
matrix. For instance,
$$
V = \left({V\atop 0} ~~ {0\atop V}\right)
$$
For $\vec A = 0$, $V = -{Ze^2\over r}$, Dirac found the solution. The energy
of the levels,
including the rest mass is given by
\beq
E(n,j) = ~mc^2~~\left[ 1 + {(Z\alpha)^2\over (n-j-1/2 + \sqrt{(j+1/2)^2 -
(Z\alpha )^2})^2}\right]^{-1/2}
\label{fortythree}
\eeq
\beq
E(n,j) \simeq m~ c^2 - {m~c^2(Z\alpha)^2\over 2 n^2}~~\left [ 1 +
{(Z\alpha)^2\over n}~~\left({1\over j+1/2} - {3\over 4n}\right)\right]
\label{fortyfour}
\eeq
where $j$ is the angular momentum. There is, however, an extra quantum
number
\beq
k = \pm \vert j + 1/2 \vert~,
\label{fortyfive}
\eeq
such that $k = + \vert j + 1/2 \vert$ if $j = \ell -1/2$ and $k  = - \vert
j + 1/2 \vert$ if $j =
\ell - 1/2$ where $\ell$ is the orbital angular momentum. However, for a
pure Coulomb potential
the two levels coincide, for instance
$$
E(2P_{1/2}) = E(2S_{1/2})~.
$$
As we said already, the Lamb Shift, an effect of quantum electrodynamics, 
leads to a violation of
this equality.

It has been shown by Grosse, Martin and Stubbe \cite{kk} that if the
potential V is attractive and
such that $\Delta V < 0$, the energy levels satisfy:
\bea
\left.
\matrix{
&&2 P _{3/2} > 2P_{1/2} > 2S_{1/2}\hfill \cr
&&3D_{5/2} > 3D_{3/2} > 3P_{3/2} > 3P_{1/2} > 3S_{1/2}}\right\} \quad{\rm
etc.}
\label{fortysix}
\eea
Eq.~(\ref{fortysix}) shows that even if the outer electron of sodium is treated relativistically, the problem of the ordering of the $D$ states remains.  So it is due to the one-electron approximation.

In the non-relativistic limit, Dirac finds:\\
1) that the electron has a gyromagnetic factor $g = 2$,\\
2) that the spin-orbit coupling is correct, including the Thomas precession.

This was a \underline{tremendous} success.

The usual method to get the non-relativistic Hamiltonian is to carry a
Foldy-Wouthuysen
transformation. The problem, with this Hamiltonian is that there are terms
like
$$
(\vec\sigma \cdot \vec L)~~{1\over r} ~~{dV\over dr}
$$
which as we said already are not lower bounded in the case of a Coulomb
potential. The trick of
Gestezy, Grosse and Thaller \cite{bb} is to consider the Pauli Hamiltonian
(without the spin-orbit
term) as the limit for $c\rightarrow\infty$ of the Dirac Hamiltonian and to
show that the resolvant
of the Dirac Hamiltonian is holomorphic in $1/c$ around $c = \infty$:
\beq
(H_c - mc^2 - Z)^{-1} = \left({(H_P-Z)^{-1}\atop 0}~~{0\atop 0}\right) +
0\left({1\over c}\right)
\label{fortyseven}
\eeq
with
\beq
H_P = {(\vec\sigma\cdot\vec\pi)^2\over 2m} + V(x)
\label{fortyeight}
\eeq
they show that
\beq
\left({1~0\atop 0~c}\right)~~(H_c = mc^2 - Z)^{-1} \left({1~0\atop
0~0}\right)^{-1}
\label{fortynine}
\eeq
can be expanded in powers of $1/c^2$.

If $E_0$ is the eigenvalue of $H_p$, with eigenvalue $f_0$, the first
correction is
\beq
E_1 = \left({\vec\sigma\cdot\vec\pi\over 2m} f_0~,~~
(V-E_0)~{\vec\sigma\cdot\vec\pi\over 2m}
f_0\right)~.
\label{fifty}
\eeq
They calculate the next one, but it is too lengthy to be written here, and
they could iterate their
procedure. $E_1$ can be rewritten as
\bea
E_1  &=& {1\over 4m^2} < f_0 \vert (V-E)~(\vec\sigma\cdot\vec\pi)^2\vert f_0
>
\nonumber \\
&+&  {1\over 4m^2} < f_0 \vert {dV\over dr}~{d\over dr}\vert f_0 > \nonumber
\\
&+&{1\over 4m^2} < f_0 \vert  {1\over r}~{dV\over dr}~(\vec\sigma\cdot
x \wedge \vec\pi )\vert
f_0 >~.
\label{fiftyone}
\eea
The last term is exactly the spin orbit coupling (for a constant magnetic
field). The second term is
not affecting the fine structure. The first term, if one disregards
higher-order terms in $A$ is
just
\beq
- < f_0 \vert {p^4\over 8 m^3}\vert f_0 >~,
\label{fiftytwo}
\eeq
a relativistic correction to the kinetic energy which is exaclty what you
would get using the
semi-relativistic kinetic energy:
\beq
E_{kin} = \sqrt{m^2+p^2} - m = {p^2\over 2m} - {p^4\over 8 m^3} +  \ldots
\label{fiftythree}
\eeq

It is an elementary exercise to check that for zero magnetic field, $E_1$
\underline{preserves the
degeneracy} \underline{ in $\ell$} for a given $j$, by a cancellation
between the first and the last
term.

All this is very nice, but what to do if you have many electrons and you
want to use the Pauli
hamiltonian? You have to decide to treat the spin-orbit terms as
perturbations, as well as spin-spin
terms which also exist, or to put a cut-off in the spin orbit interaction.
For one electron
\underline{one can show} that this cut-off should be at $r_0$ such that
$\vert V(r_0)\vert\simeq
2m$. One could decide to use the same rule for many electrons.

The other very remarkable property of the Dirac equation is the occurrence
of \underline{negative
energy} \underline{ states}. This seemed to be a terrible handicap for the
Dirac equation but turned
out to lead to the fantastic discovery of antiparticles and antimatter.

The trick that Dirac invented was to assume that all negative energy states
were filled, and then,
because of the Pauli exclusion principle, they could not be  occupied by
other electrons, and became
harmless. However, if one had enough energy to extract one of these
electrons from the sea, one
would see an electron moving normally and a hole which would behave like a
positively charged
particle. At the beginning, Dirac believed that this positively charged
particle was the proton.
However, Oppenheimer pointed out that in that case the proton would
annihilate with the electron
in hydrogen to produce photons (in fact, many years later, Martin Deutsch
manufactured positronium,
a system made of an electron and an antielectron which do annihilate rapidly
into photons). Then
Hermann Weyl pointed out that the field theory of electrons should be
completely symmetrical between
particles and antiparticles. As you know the antielectron, called positron,
was discovered in 1932
by Anderson. Positrons are now very easily produced in large amounts and
used in particular to
study electron-positron collisions in machines, the last one being LEP in
Geneva.

\section{Spin of  Particles Other than the Electron}

You might think that the spin of the proton was obtained first from an
experiment of the
Stern-Gerlach type. Indeed, Stern did such an experiment, but only later.
The first indication that
the proton had spin 1/2 came from the observation of an anomaly in the
specific heat of the
molecular hydrogen around 1927.

If protons have spin 1/2, the two protons inside the hydrogen can have a
spin 1 and hence a
symmetric wave function -- this is called \underline{ortho}hydrogen -- or a
spin 0, with an
antisymmetric wave function, which is called \underline{para}hydrogen.
 These two protons are bound by a potential which is produced by the
electrons (in the framework of
the Born-Oppenheimer approximation), and they have rotation levels
(vibrations also exist but are
much higher).

\underline{Ortho}hydrogen has only \underline{odd} angular momentum rotation
levels because of the
Pauli principle for protons, while
\underline{para}hydrogen has only even rotation levels. Taking this into
account in counting the
degrees of freedom of hydrogen, with a ratio 3:1 of ortho 
\underline{para}hydrogen at room
temperature all anomalies in the specific heat are removed. I cannot
reproduce the rather intricate
details of the argument which are given in Tomonaga's book.

Then Otto Stern and Extermann did a Stern-Gerlach experiment on the proton
in 1933, confirming of
course the spin 1/2 but they also measured the magnetic moment of the
proton. There are two stories
about this. Jensen (the German physicist who got the Nobel prize with Maria
Goeppert-Mayer for the nuclear 
shell model) told Tomonaga that Pauli was  visiting Stern's laboratory and,
when Stern explained
that he wanted to measure the magnetic moment of the proton, Pauli said it
was useless because the
gyromagnetic factor would be 2, like for the electron. The second story, 
Gian Carlo Wick (who
incidentally, spent the last years of his life here, in Turin) told me. Otto
Stern entered a seminar
room in Leipzig, where all the elite of physics was present, Heisenberg,
Oppenheimer, etc., told he
had measured the magnetic momentum of the proton and asked the people
present to indicate on a sheet
of paper which value it had, with their signature. Most said it would be one
proton Bohr magneton,
i.e., $g = 2$. Then Stern announced that he had found 2.5. Bohr magnetons!
Now we know that it is
2.79 Bohr magnetons.

This discrepancy is the only reason why one might doubt that the proton, in
spite of its spin 1/2,
is a good Dirac particle with an associated antiparticle, the antiproton,
and design a special
experiment in 1956 to prove the existence of the antiproton, and give the
Nobel prize to Segr\'e and
Chamberlain. Yet, already then, it was known that the gyromagnetic factor of
the electron was not
exactly 2, but
\beq
g_e = 2 + {\alpha\over\pi} + \ldots~, \quad {\rm with} ~\alpha = {1\over
137}~,
\label{fiftyfour}
\eeq
the first correction being due to J. Schwinger. Higher corrections have been
calculated by our
friends Kinoshita $\ldots$, Remiddi, etc $\ldots$ , and very recently, De
Rafael, Knecht, Nyffeler and
Perrotet  \cite{lll}.

Now let me present an argument ``explaining" the anomalous gyromagnetic
ratio of the proton. I hope
that my QCD friends, like E. De Rafael, H. Leutwyler, etc., will forgive me.

You know that the proton and the neutron are made of quarks:
$$
\matrix{
{\rm proton}~ =  u~~u~~d \hfill\cr
{\rm neutron} = u~~d~~d \hfill\cr
{\rm where} ~u ~{\rm  has~ a~ charge} ~+2/3 \cr
~~~~~~~~d~ {\rm has ~a~ charge} ~-1/3~.
}
$$

In a very \underline{na\"\i ve} view, the binding energy of the quark could
be \underline{small}.
Then
\beq
m_n = m_d = {m_p\over 3} \simeq {m_n\over 3} \simeq 312 ~{\rm MeV}
\label{fiftyfive}
\eeq
(The QCD masses of the $u$ and $d$ quarks are a few MeV.)

Assume now that the gyromagnetic ratio of $u$ and $d$ is 2, like good Dirac
particles. Then you get
\bea
\mu_p &=& 3 ~{\rm Bohr~proton~magnetons}\left({{\rm experiment}\atop 
~~2.79}\right)
\nonumber \\
&& \nonumber \\
\mu_n &=& -2~ {\rm Bohr~proton~magnetons}\left({{\rm experiment}\atop 
-1.91}\right)
\label{fiftysix}
\eea
Since questions were asked about that, we give the proof:
$$
{\rm proton} = u~u~d
$$

The colour wave function of the three quarks is a singlet completely
antisymmetric. Hence
\underline{in the ground state} the $u~u$ system has a space and spin wave
functions which are
\underline{symmetric}, i.e., spin 1. We construct a proton with $s_Z = +1/2$
using Clebsch-Gordan
coefficients:
\beq
\bigg\vert ~{1\over 2}, {1\over 2} > = \sqrt{{2\over 3}}~~\bigg\vert
~1~1>_{uu}~~\bigg\vert {1\over
2} - {1\over 2}  > _d +
\sqrt{{1\over 3}}~
\bigg\vert~ 1~0>_{uu}~~\bigg\vert {1\over 2} + {1\over 2} >_d
\label{fiftyseven}
\eeq
Hence the magnetic moment is
$$
\left(~{2\over 3}~\bigg[~ 2\times {2\over 3} + {1\over 3} ~\bigg] + {1\over
3}~\bigg[~0 - {1\over
3}~\bigg]\right)~~{ \hllap \over 2 m_u c} = 3 \left({e\hllap\over
2m_pc}\right)
$$
if $m_u = m_d = {m_p\over 3}$.

In the neutron case, it is the $dd$ pair which has spin1 and we get
$$
\left(~{2\over 3}~\bigg[~ 2\times \left({-1\over 3}\right) - {2\over 3}
~\bigg] + {1\over 3}~
\bigg[~+ {2\over 3}~\bigg]\right)~~3~{e \hllap \over 2 m_p c} = -2
\left({e\hllap\over 2m_pc}\right)~.
$$

You can apply this to all baryons made of $u, d, s$ quarks. The strange
quark must be
\underline{heavier} than the $u-d$ quarks. From the masses of the $\Lambda$
and $\Sigma$ hyperons we
see that the strange quark must be about 200 MeV heavier, which gives
\beq
\frac{m_{u,d}}{m_s} = 0.625 ~.
\label{fiftysevenforgot}
\eeq
With this value, we get the following table, which is in my opinion 
remarkable (remember that \underline{just getting the right  signs is not 
trivial}).  $m_u/m_s = 1$ is much less good.  All the particles listed 
have a spin 1/2 except the $\Omega^-$.

Notice that there is one prediction which is mass independent which is:
\beq
\mu_{\Omega^-} (spin 3/2) = 3 \mu_\Lambda~~\left( {spin 1/2 \atop
u~d~singlet}\right)~,
\label{fiftyeight}
\eeq
which is approximately satisfyied by experiment.
{\small
\vspace*{0.5cm}
\begin{center}
Table 2\vglue.3cm
\begin{tabular}{|l |l| c| c | }
\hline
&      & $\mu /\mu_P$  & \\
& $\exp$ & Quark Model & Quark Model \\
&      & $m_u/m_s = 1$ & $m_u/m_s = 0.625$ \\
\hline
N & ~~-0.68 & -0.67 & -0.67 \\
\hline
$\Lambda$ & ~~-0.22 & -0.333 & 0.207 \\
         & $\pm$ 0.02 & & \\
\hline
$\Sigma^+$ & ~~0.85 & 1 & 0.958 \\
& $\pm$ 0.01 & & \\
\hline
$\Sigma^-$ & ~~-0.50 & -0.333 & -0.37 \\
& $\pm$ 0.059 & &  \\
\hline
$\Xi^-$ & ~~-0.24 & -0.333 & -0.17 \\
        & $\pm$ 0.01 & & \\
\hline
$\Xi^0$  & ~~-0.45 & -0.666 & -0.50 \\
& $\pm$ 0.05 && \\
\hline
$\Omega^-$ & ~~-0.72 & -1 & -0.625 \\
& $\pm$ 0.07 & & \\
\hline
\end{tabular}
\end{center}}

Finally, since the proton has a magnetic moment, it has a spin-spin
interaction with the electron,
which gives rise to an hyperfine splitting. The ``rule" is that it is given
by
\beq
W = {a\over 2}~\bigg [ F(F+1) - I(I+1) - J(J+1) \bigg]
\label{fiftynine}
\eeq
$F$ = total angular momentum \\
$I$ = angular momentum of the nucleus \\
$J$ = angular momentum of the electron
\beq
a = {16\pi\over 3} ~g_e~~g_{Nucl} ~~\mu_{BN}~~\mu_{Be}~~\vert\psi (0)\vert^2
\label{sixty}
\eeq
where $\psi(0)$ is the \underline{Schr\"odinger} wave function at the origin
(the Dirac wave
function at the origin is infinite!). This is not trivial! (worse than the
spin-orbit coupling!).
For Hydrogen this gives the famous 21 cm line, produced by an excited state
which has a lifetime of
$10^7$ years.

Of course the year 1932 is also the year of the discovery of the neutron by
Frederic and Ir\`ene
Joliot-Curie and James Chadwick. The Joliots had seen a mysterious
penetrating neutral particle.
According to Gian-Carlo Wick, when Majorana read the ``note aux Comptes
Rendus" of the Joliots, he
exclaimed: ``Stronzi! Non hanno capito ch\'e \`e il neutrone!"~~``Idiots!
They have not understood
that it is the neutron!"

Chadwick really demonstrated that it is a neutron by a kind of billiard
experiment: when a neutron
hits a proton, the recoiling proton and the deflected neutron go at right
angle because they have
equal masses.

That the neutron has also spin 1/2 it is not a surprise. It was shown by J.
Chadwick and M.
Goldhaber (still alive!) in 1934 that the deuteron was made of a proton and
a neutron by
dissociating it with a gamma ray:
$$
\gamma + d \rightarrow n + p
$$
They showed that the masses of the proton and the neutron were very close,
the neutron being
slightly heavier. The spin and statistics of the neutron were obtained from
the spin and statistics
of the deuteron from the band spectrum of the deuteron molecule. The
deuteron has spin 1. The
neutron has hence a half-integer spin, 1/2 or 3/2.  3/2 was very unlikely
and not so long ago (since
nuclear reactors producing neutrons exist!) Stern-Gerlach experiments have
been performed and the
spin found to be 1/2 with a magnetic moment -1.93 $\mu_{Bn}$. The magnetic
moment was first deduced
from the magnetic moment of the deuteron, using an extremely plausible
theoretical model of the
deuteron.

Previously, Rasetti, who died very old only a few months ago, here in Turin
I believe, had proved
that the spin of $N_{14}$ which has charge 7 is zero from the band
structure of the Nitrogen
molecule. This demonstrated that it was made of an equal number of protons
and neutrons.

\section{Spin and Statistics and the Stability of the World}

We have seen that electrons have spin 1/2 and satisfy the Pauli principle.
They are usually said to
satisfy Fermi (or Fermi-Dirac) statistics. Why Fermi? Probably because Fermi
studied the free
electron gas which is, to a good approximation, what you find in a
conductor. The energy levels are
filled up to a certain level, the Fermi level.

There is another type of statistics: Bose statistics. Bosons are integer
spin particles with a wave
function which is completely symmetrical. They were invented in 1927 by S.M.
Bose, who sent a letter
to Einstein, who immediately understood the importance of Bose's discovery
and developed it so that
now people speak of ``Bose-Einstein" particles.

Concerning Dirac particles with spin 1/2 it is intuitively clear that if one
does not want to have
a catastrophy they must satisfy the Pauli principle. Otherwise Dirac's
original argument with the
filling of the energy levels breaks down. This remains true in the modern
field theoretical picture
for electrons and positrons. I have neither the time nor the ability to
explain this further.

As for integer spin particles, spin 0 like the pion, or 1 like the
intermediate vector bosons
$W^\pm$ and $Z_0$, they must satisfy Bose statistics as shown very early by
Pauli and Weisskopf (who
just died!). Bosons, except for the motion due to uncertainty relations, can
-- crudely speaking --
be all at the same place. This is realized in boson condensates, the
observation of which has been
rewarded by the 2001 Nobel prize.

In our world, with one time and three space dimensions there is only room
for these two kinds of
statistics. In two spatial dimensions there is room for other statistics.
This is important for the
quantum Hall effect but we shall not speak about that here.

If the world exists as it is, it is essentially because electrons have spin
1/2 and satisfy the
Pauli principle. The late Weisskopf told me that if he puts his hand on a
table and the hand does
not go through the table it is because the electrons of the hand cannot be
at the same place as the
electrons of the table. If we start with atoms satisfying the Schr\"odinger
equation, heavy atoms
are well described by the Thomas-Fermi model which gives \cite{mm}
$$
E \sim -0.77 \ldots Z^2~~N^{1/3}~\alpha^2~m~,
$$
which has been shown to be asymptotically exact by Lieb~\cite{nn}
for $Z\rightarrow\infty$
 (this limit is of course purely \underline{mathematical}  since the
Schr\"odinger equation is no
longer valid because of relativistic effects) and furthermore is presumably
a \underline{lower
limit} of the binding energy of the atom according to Lieb and Thirring
\cite{oo}, modulo a certain
unproved technical assumption on bound states in potentials. The two
ingredients of the Thomas-Fermi
model are:
\begin{itemize}
\item[i)] the electrostatic repulsion between electrons and the electric
attraction of the nucleus;
\item[ii)] a crude form of the Pauli principle: locally the electrons are
considered as free in a
potential well and their density is determined by the Fermi level.
\end{itemize}

Dyson and Lennard \cite{pp} have gone further than that and established what
they call the
``stability of matter" in a non-relativistic world where only electric
forces are taken into
account, the fact that the binding energy of an assembly of $N$ atoms has an
energy which behaves
like $-CN$ and a minimum volume which behaves like $C'N$. Only the Fermi
statistics of the electrons
is essential. The spin of the nuclei can be either integer or half integer
(for instance liquid
Helium has a volume proportional to $N$). What is remarkable in the work of
Lieb-Thirring \cite{oo}
is that they succeed in getting a constant $C$ which is not more than twice
a realistic value.

Things change if either all particles satisfy Bose statistics or if instead
of electric forces we
have gravitational forces.  Table 3 \cite{qq} gives a summary, for the
\underline{non-relativistic}
situation (Schr\"odinger equation).
{\small
\vspace*{0.5cm}
\begin{center}
Table 3\vglue.3cm
\begin{tabular}{|l |l| c| c | }
\hline
Nature of the Forces & Statistics      &Volume  & Energy \\
\hline
electric, with & Fermi & N & -N \\
\underline{finite mass} &  &  &  \\
positive and negative   & Bose & $N^{-3/5}$ & -N$^{7/5}$ \\
particles &&&\\
\hline
 & Fermi & N$^{-1}$ & -N$^{7/3}$ \\
gravitational&  & & \\
& Bose & N$^{-3}$ & -N$^3$ \\
\hline
\end{tabular}
\end{center}}

Now, what happens if we put in relativistic effects? We already know that
the Dirac equation with a
point nucleus of charge $Z$ and an electron breaks down if
$$
Z\alpha = Z~{e^2\over \hllap c} = 1~, ~~~{\rm i.e.,}~~~Z \sim 137~.
$$
the Klein-Gordon equation breaks down for $Z\alpha = 1/2$ and the ``Herbst"
or Salpeter equation,
$$
\left(\sqrt{p^2+m^2} - {Z\alpha\over r} - E\right)~~\psi = 0
$$
breaks down for $Z\alpha = 2/\pi$. This occurs also for many-body systems.

Take the case of an assembly of $N$ particles with all equal masses (for
simplicity) $m$ \cite{rr}.
Call $E(N,m)$ the non-relativistic energy of this system:
$$
\left[ \Sigma ~ {p^2_i\over 2m} + V(x_1\ldots x_n)\right]~~\psi =
E(N,m)~\psi
$$
In a \underline{semi-relativistic} treatment, one replaces
$m + p_i^2/ m$ by $\sqrt{p^2_i + m^2}$. To get the ground state energy one
has to minimize the
expectation value of
$$
H_{Rel} = \sum \sqrt{p^2_i + m^2} + V(n_1\ldots n_n)
$$
but
$$
\sqrt {p^2_i + m^2} \geq {1\over 2}~~\left[ M + {p^2+m^2\over M}\right]
$$
So
$$
\langle H_{Rel}\rangle \leq inf_M \left[ {n\over 2} ~~(M + {m^2\over M}) + E(N,M)
\right]
$$
So, if
$$
E(N,m) \sim -C m N^\alpha \quad {\rm for~large} \quad N
$$
(it comes from homogeneity), we have to minimize
$$
{N\over 2}~~\left( M + {m^2\over M}\right) - C~M~N^\alpha~.
$$

We see immediately that if $\alpha > 1$ there is \underline{ no minimum} for
sufficiently large $N$
and the system collapses. This is the case of an electric system with Bose
statistics or of any
gravitational system. The above equation gives an upper bound of the
Chandrasekhar limit for a
neutron star which is not unreasonable and also a limit on the mass of a
boson sate \cite{ss}.

In the case of fermions, or Fermion-Boson, the occurrence of the collapse
will depend on the value
of $C$, which is itself linked to the strength of the coupling (this was
already the case for one or
two particles).

The (temporary) existence of our world depends on a careful balance between
masses and interactions
of its constituents.


\newpage
\begin{thebibliography}{99}
\bibitem{aaa} B. Baumgartner, H. Grosse and A. Martin, \PL {\bf B284} (1992) 347.

\bibitem{bb} F. Gesztesy, H. Grosse and B. Thaller, \PRL {\bf 50} (1983)
625.

\bibitem{cc} A. Martin and J. Stubbe, {\it Europhysics Letters} {\bf 14} (1991) 287.

\bibitem{dd} A. Martin and J. Stubbe, \NP {\bf B367} (1991) 158.

\bibitem{ee} A. Aspect, \PR {\bf 14D} (1976) 1944.

\bibitem{ff} J.-M. Richard, J. Fr\"ohlich, G.M. Graf and H. Seifert, \PRL {\bf 71} (1993) 1332.

\bibitem{ggg} G.B. Lesovik, T. Martin and G. Blatter, {\it European Physical Journal B} {\bf 24} (2001) 287.

\bibitem{hh} J.S. Bell, Journal de Physique, Colloque C2, Supplement to No. 3 Tome 42 (1982) 820.

\bibitem{jj} B.L. Van der Waerden, Die gruppentheoretische Methode in der Quantenmechanik, Berlin:  Springer, 1932.
\bibitem{kk} H. Grosse, A. Martin and J. Stubbe, \PL {\bf 284} (1992) 347.

\bibitem{lll} M. Knecht, A. Nyffeler, M. Perrottet and E. De Rafael, \PRL {\bf 88} (2002) 071802.

\bibitem{mm} See, for instance W. Thirring, Quanten Mechanik grosser systeme, Springer-Verlag, Wien (1980) p. 197.

\bibitem{nn} E.H. Lieb, \PL {\bf 70A} (1979) 444.

\bibitem{oo} E.H. Lieb and W. Thirring, \PRL {\bf 35} (1975) 687.

\bibitem{pp} J.F. Dyson and A. Lennard, {\it J. Math. phys.} {\bf 8} (1967) 423 and {\it J. Math. Phys.} {\bf 9} (1968) 698.

\bibitem{qq} W. Thirring, Ref. [12] p. 19;\\
J.G. Conlon, IX Int. Congress on Mathematical Physics, B. Simon, A. Truman, I.M. Davis, Eds., Adam Hilger, Bristol (1988), p. 48.

\bibitem{rr} A. Martin, \PL {\bf 214} (1988) 561.

\bibitem{ss} A. Martin and S.M. Roy, \PL {\bf B233} (1989) 407;\\
J.C. Raynal, S.M. Roy, V. Singh, A. Martin and J. Stubbe, \PL {\bf B320} (1994) 105.
\end{thebibliography}
\end{document}